\documentclass[nofootinbib,amsmath,amssymb,prd,twocolumn,10pt]{revtex4-2}

\usepackage{mathrsfs}
\usepackage[x11names]{xcolor}
\usepackage{graphicx}
\usepackage{dcolumn}
\usepackage{bm}
\usepackage[pdftex]{hyperref}
\usepackage{soul}
\usepackage{float}
\usepackage{orcidlink}
\usepackage[skip=5pt, indent=10pt]{parskip}
\usepackage{tabularray}
\usepackage[normalem]{ulem}

\hypersetup{pdftitle={Neutrino Oscillations as a Probe of Macrorealism},
pdfsubject={Neutrino Oscillations as a Probe of Macrorealism},
pdfauthor={Groth, Ioannou-Nikolaides, Koskinen \& Ahlers},
pdfstartview={FitH},
colorlinks=true,
bookmarksopen=false,
bookmarksnumbered=false,
bookmarksopenlevel=0,
linkcolor=Blue1!60!black,
citecolor=Green1!50!black,
urlcolor=Blue1!70!black
}

\begin{document}

\title{Neutrino Oscillations as a Probe of Macrorealism}

\author{Kathrine Mørch Groth\,\orcidlink{0000-0002-1581-9049}}
\email{kathrine.groth@nbi.ku.dk}
\affiliation{Niels Bohr Institute, University of Copenhagen, Blegdamsvej 17, 2100 Copenhagen, Denmark}

\author{Johann Ioannou-Nikolaides\,\orcidlink{0000-0003-2486-1588}}
\affiliation{Niels Bohr Institute, University of Copenhagen, Blegdamsvej 17, 2100 Copenhagen, Denmark}

\author{D.~Jason Koskinen\,\orcidlink{0000-0003-0709-5631}}
\affiliation{Niels Bohr Institute, University of Copenhagen, Blegdamsvej 17, 2100 Copenhagen, Denmark}

\author{Markus Ahlers\,\orcidlink{0000-0003-0709-5631}}
\affiliation{Niels Bohr Institute, University of Copenhagen, Blegdamsvej 17, 2100 Copenhagen, Denmark}

\begin{abstract}
The correlations between successive measurements of a quantum system can violate a family of Leggett-Garg Inequalities (LGIs) that are analogous to the violation of Bell's inequalities of measurements performed on spatially separated quantum systems. These LGIs follow from a {\it macrorealistic} point of view, imposing that a classical system is at all times in a definite state and that a measurement can, at least in principle, leave this state undisturbed. Violations of LGIs can be probed by neutrino flavour oscillations if the correlators of consecutive flavour measurements are approximately stationary. We discuss here several improvements of the methodology used in previous analyses based on accelerator and reactor neutrino data. We argue that the strong claims of LGI violations made in previous studies are based on an unsuitable modelling of macrorealistic systems in statistical hypothesis tests. We illustrate our improved methodology via the example of the MINOS muon-neutrino survival data, where we find revised statistical evidence for violations of LGIs at the $(2-3)\sigma$ level, depending on macrorealistic background models.
\end{abstract}

\maketitle

\section{Introduction}

The statistical properties of measurements performed on a quantum system defy our probabilistic intuition from a macroscopic, {\it i.e.}~classical, point of view. One well-known example is the violation of Bell's inequalities in entangled quantum systems~\cite{Bell:1964kc}, which challenges the probabilistic interpretation of quantum mechanics as a local theory of hidden variables. While Bell's inequalities relate simultaneous measurements in spatially separated systems, {\it Leggett \& Garg}~\cite{Leggett:1985zz} derived analogous inequalities that relate successive measurements performed on a local system. These Leggett-Garg inequalities (LGIs) follow from three {\it macrorealistic} postulates that: {\it i)} a system is at any time in a definite state of its observables ({\it macroscopic realism per se}), that {\it ii)} a measurement on this system can, in principle, be performed without disturbance ({\it noninvasive measurability}), and that {\it iii)} the properties of the system are exclusively determined by its initial condition ({\it induction})~\cite{Leggett:2002}. These three postulates appear to be reasonable from a classical perspective and they define our notation of the term {\it macrorealism} in the broader sense; for a review see Ref.~\cite{Emary:2013wfl}.

It has been argued that the violation of LGIs -- and therefore macrorealism in the broader sense -- can be observed in neutrino flavour oscillations~\cite{Gangopadhyay:2013aha,Gangopadhyay:2017nsn}. This first appears counter-intuitive since neutrinos are only weakly interacting and therefore a successive measurement of a neutrino system is experimentally unfeasible. However, if the correlation of neutrino flavours measured at different times is stationary and therefore only depends on the time difference, a sequence of measurements can be represented by neutrino flavour survival and transition probabilities that are inferred from a large ensemble of equally prepared neutrinos, {\it e.g.}~a beam of muon neutrinos from pion decays. In this way, previous work has studied violations of LGIs in accelerator neutrino beams from MINOS~\cite{Formaggio:2016cuh} and NO$\nu$A~\cite{Barrios:2023yub} as well as reactor neutrinos from Daya Bay~\cite{Fu:2017hky}, RENO~\cite{Barrios:2023yub}, and KamLAND~\cite{Wang:2022tnr}; see also Refs.~\cite{Song:2018bma,Naikoo:2019eec,Shafaq:2021lju}.

The standard LGIs used in previous analyses of neutrino data are based on Leggett-Garg {\it strings} (called $K_n$ in the following; see Ref.~\cite{Emary:2013wfl}) which can be constructed from a combination of $n$ neutrino flavour measurements at oscillation phases ($\phi$) related to the neutrino baseline ($L$) and energy ($E$) as $\phi \propto L/E$ with the requirement that the sum over the first $n-1$ phases matches the $n$th phase within some tolerance. A test statistic for the departure from macrorealism can then be defined as the fraction of violations of LGIs inferred from all of these matching phase combinations. We discuss in this paper a generalization of this method, that allows us to probe LGIs between arbitrary sequences of data with matching phase sums and that always yields the strongest violation of LGIs for these matching sequences. 

A statistical hypothesis test of LGIs needs to be able to quantify in what way the observed neutrino flavour correlations are incompatible with the predictions from macrorealism. Previous work has approached this question by introducing a {\it classical} LG string ($K^{\rm C}_n$) and testing its violation of LGIs on re-sampled oscillation data~\cite{Formaggio:2016cuh,Fu:2017hky}, rather than simulating pseudo-data following a macrorealistic background model; other studies have neglected to study the behaviour of background models~\cite{Wang:2022tnr, Barrios:2023yub}. We argue here that this approach does not yield reliable results and the significance of the violation of LGIs reported in previous studies should be scrutinized. In particular, the {\it classical} LG string $K^{\rm C}_n$ when applied to oscillation data will -- by construction -- only violate LGIs if the flavour correlations become unphysical, related to uncertainties of experimental data. In this paper, we provide an alternative statistical method that offers a more robust estimate of the significance. 

The paper is organized as follows: in Section~\ref{ch1} we first review the general form of LGIs based on LG strings and their violation in quantum-mechanical systems. We then introduce an improved method for identifying LGIs for stationary correlators in neutrino flavour oscillation analyses in Section~\ref{ch2}. In Section~\ref{ch3} we discuss our revised method for testing macrorealism in neutrino oscillations based on pseudo-data, that is robust against unphysical flavour correlators, $|\mathcal{C}|>1$, unlike previously proposed methods. We then apply our method to MINOS/MINOS+ data in Section~\ref{ch4} in order to demonstrate the differences and improvements compared to previous studies. We discuss our results in Section~\ref{ch5} before concluding in Section~\ref{ch6}. Throughout the paper we are working in natural units where $c=\hbar=1$ and boldface quantities indicate vectors.

\section{Leggett-Garg Inequalities}\label{ch1}

In the following, the quantity $Q(t)$ represents a {\it dichotomic} property of a system evolving in time $t$ with values normalized to $\pm1$. For instance, we can assign $Q(t)=1$ for the case that a neutrino is in the flavour state $\nu_\mu$ and $Q(t)=-1$ for all other flavour states, {\it i.e.}~$\nu_e$ or $\nu_\tau$. We will use the abbreviation $Q_i = Q(t_i)$ for different observation times $t_i$ in the following. The expected correlation of measurements at different times is expressed by the correlators $\mathcal{C}_{ij} \equiv \langle Q_iQ_j\rangle$ which are bounded as $|\mathcal{C}_{ij}| \leq 1$ with $\mathcal{C}_{ij} = \pm1$ representing full correlation and anti-correlation, respectively. The theoretical prediction of the expectation value $\langle\cdot\rangle$ is fundamentally different in quantum mechanics (QM) and macrorealism (MR).

Leggett-Garg inequalities express relations between the correlators $\mathcal{C}_{ij}$ under the assumption of MR~\cite{Leggett:1985zz}. From a macrorealistic point of view, the expectation values $\langle\cdot\rangle$ are understood as ensemble averages following from underlying joint probability densities $\rho(Q_i,Q_j)$ of measuring the outcomes $Q_i$ and $Q_j$ at times $t_i$ and $t_j$, respectively~\cite{Leggett:1985zz}. More explicitly, the macrorealistic correlator is expected to take the form:
\begin{equation}\label{eq:CMR}
{\mathcal{C}}_{ij} \equiv \sum_{Q_i,Q_j}Q_iQ_j\rho(Q_i,Q_j)\qquad \text{(in MR)}\,.
\end{equation}
Note that MR requires that the correlator in Eq.~(\ref{eq:CMR}) is symmetric under the exchange of measurement times, $\mathcal{C}_{ij} = \mathcal{C}_{ji}$. One can now define $n$-measurement Leggett-Garg strings $K_n$ with $n\geq 3$ as~\cite{Emary:2013wfl}:
\begin{equation}\label{eq:Kn}
K_n \equiv \sum_{i=1}^{n-1}\mathcal{C}_{i(i+1)}- \mathcal{C}_{1n}\,.
\end{equation}
Using the macrorealistic correlator in Eq.~(\ref{eq:CMR}) it can be shown that these LG strings obey the LG inequalities $K_n \leq n-2$ (see Appendix~\ref{appA}).

Conversely to a classical system, in a quantum system the correlation of two consecutive measurements can be expressed as: 
\begin{equation}\label{eq:CQM}
\mathcal{C}_{ij} \equiv \frac{1}{2}\langle \lbrace \hat Q_i, \hat Q_j\rbrace\rangle\qquad \text{(in QM)}\,,
\end{equation}
where $\hat Q_i$ and $\hat Q_j$ are understood as Heisenberg operators and the expectation value $\langle\cdot\rangle$ is now in terms of a trace over basis states weighted by the density matrix (see Appendix~\ref{appB}). Again, the correlator in Eq.~(\ref{eq:CQM}) is symmetric, $\mathcal{C}_{ij} = \mathcal{C}_{ji}$.

\begin{figure*}[t]
\includegraphics[width=\linewidth]{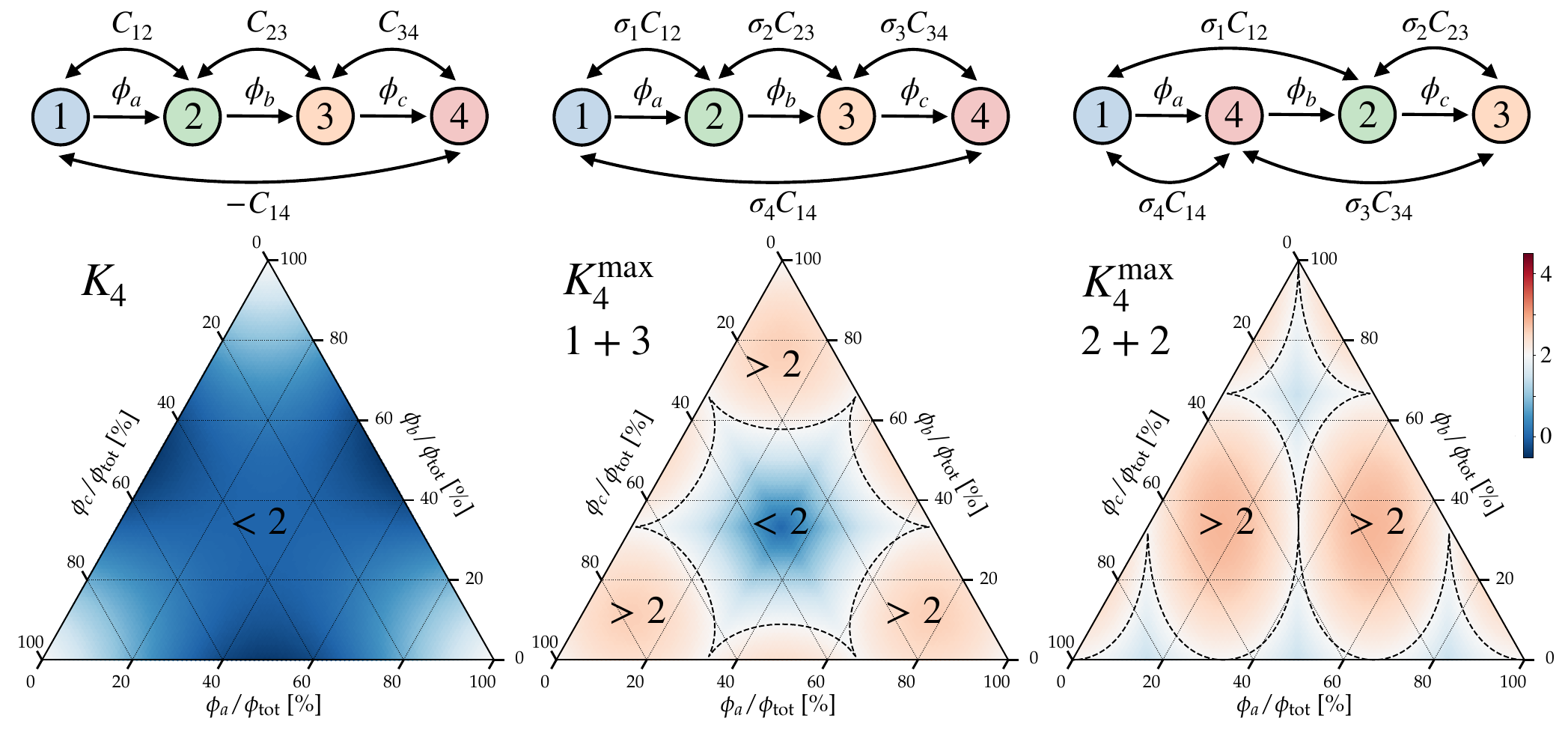}
\caption[]{Illustration of LG strings and their violation of LGIs in QM. We consider $n=4$ consecutive measurements (see top panels) with phase steps $\phi_a$, $\phi_b$, and $\phi_c$ and assume that the correlators follow $\mathcal{C}_{ij} = \cos(\phi_i-\phi_j)$ for maximal two-level oscillations in QM. The total phase $\phi_{\rm tot} = \phi_a+\phi_b+\phi_c$ is fixed to $\phi_{\rm tot}=3\pi/2$, comparable to the phase range covered by the MINOS/MINOS+ data. The ternary plots indicate the relative combinations of $\phi_a$, $\phi_b$, and $\phi_c$ where we expect that the corresponding LG strings violate (obey) the LGIs indicated by the red-shaded (blue-shaded) regions. The left panel shows $K_4$ as in Eq.~(\ref{eq:Kn}) which does not violate the LGI $K_4\leq 2$ (nor $-2\leq K_4$) for this example. The center (right) panel shows $K_4^{\rm max}$ in Eq.~(\ref{eq:Knmax}) for two different orders of measurements (``$1+3$'' or ``$2+2$'') as indicated in the top panel.}\label{fig1}
\end{figure*}

For instance, a stationary quantum system that maximally oscillates between the eigenvalues $Q=\pm1$ has correlators $\mathcal{C}_{ij} = \cos(\phi_i-\phi_j)$ where the oscillation phase $\phi$ depends on the oscillation frequency and time. If we consider a fixed phase step between consecutive measurements, $\phi_{i+1}-\phi_i = \Delta\phi$, the LG strings in Eq.~(\ref{eq:Kn}) become $K_n = (n-1)\cos\Delta\phi - \cos((n-1)\Delta\phi)$. These can be shown to become maximal at $\Delta \phi = \pi/n$ with $K_n(\pi/n) = n\cos(\pi/n)$ and violate the LGIs for any $n\geq3$; see Appendix~\ref{appB}.

The LG string in Eq.~(\ref{eq:Kn}) is not the only (and not necessarily the optimal) quantity for a test of MR between $n$ measurements. Firstly, the measurement times $t_i$ that appear in the LG string do not need to be time-ordered (or even different), as was assumed in previous analyses of the violation of macroscopic realism in neutrino oscillations~\cite{Formaggio:2016cuh,Fu:2017hky,Barrios:2023yub}. And secondly, the expression in Eq.~(\ref{eq:Kn}) can be generalized by arbitrarily redefining the observables as $Q_i \to -Q_i$, which introduces a sign flip in the corresponding correlators. As we show in Appendix~\ref{appA}, this leads to a generalization of the LG string that can be written in the form:
\begin{equation}\label{eq:Knsigma}
K_n({\boldsymbol\sigma}) \equiv \sum_{i=1}^{n}\sigma_i\mathcal{C}_{i(i+1)}\,,
\end{equation}
where we define $t_{n+1} \equiv t_1$ for later convenience and introduce the $n$ sign assignments $\sigma_i = \pm1$, which can take any combination subject to the condition:
\begin{equation}\label{eq:sigmacon}
\prod_{i=1}^n \sigma_i = -1\,.
\end{equation}
This results in $2^{n-1}$ independent LG strings (\ref{eq:Knsigma}), all subject to the inequality $K_n({\boldsymbol\sigma}) \leq n-2$. 

The original LG string in (\ref{eq:Kn}) corresponds to the assignment $\sigma_i = 1$ for all $i < n$ and $\sigma_n = -1$. For even $n$ we can also invert any sign vector, $\boldsymbol\sigma\to-\boldsymbol\sigma$, leading to an overall sign flip $K_n \to - K_n$ of the LG string and the two-sided inequality $-n+2\leq K_n \leq n-2$. For odd $n$ we can only give the lower bound $-n \leq K_n$, which follows from $|K_n|\leq n$.

The freedom of choosing $\boldsymbol\sigma$ in Eq.~(\ref{eq:Knsigma}) allows us to define the {\it optimal} LG string for a test of MR as the maximum:
\begin{equation}\label{eq:Knmax}
K_n^{\rm max} \equiv \max_{\boldsymbol\sigma} K_n(\boldsymbol\sigma)\,.
\end{equation} 
From Eq.~(\ref{eq:Knsigma}) it is clear that the sign assignment $\sigma_i = \text{sign}\,\mathcal{C}_{i(i+1)}$ maximizes $K_n(\boldsymbol\sigma)$ as long as this combination obeys the condition in Eq.~(\ref{eq:sigmacon}). If this is not the case, we can find the maximum by alternating the sign for the smallest member of the set $\lbrace|\mathcal{C}_{i(i+1)}|\rbrace$.

As an illustration, Fig.~\ref{fig1} shows the example of a LG string with $n=4$ and its violation of the LGI in maximally oscillating two-level quantum systems. We consider four measurements $Q_i$ that are distributed with relative phase steps $\phi_a$, $\phi_b$, and $\phi_c$ with $\phi_a+\phi_b+\phi_c = 3\pi/2$. The three different columns show different sign assignments and different time orders of measurements. The left column assumes that the measurements appear time-ordered in the LG string as indicated in the graph. The ternary diagram shows the value of the conventional LG string $K_4$ in Eq.~(\ref{eq:Kn}); none of the phase configurations violate the LGI $K_4\leq2$ (nor the lower bound $-2\leq K_4$).

In the center column of Fig.~\ref{fig1} we use the same time-ordered measurements, but now the optimized LG string of Eq.~(\ref{eq:Knmax}). With the appropriate choice of $\boldsymbol\sigma$, we now find regions where the LGI is violated (red-shaded regions). Finally, the last column of Fig.~\ref{fig1} shows an example where the measurements $Q_i$ are not time-ordered, but follow the sequence shown in the top graph. In the previous two cases, the relative phase differences $\Delta\phi_i \equiv \phi_{i+1}-\phi_i$ were positive for $i\leq 4$ and negative for $i=4$ (``$1+3$''). Now, we have two positive ($i=1$ and $2$) and two negative ($i=3$ and $4$) phase differences between observations (``$2+2$''). Note that the corresponding LG strings of Eq.~(\ref{eq:Knmax}) now finds violations of LGIs that were not visible under the previous time order (left \& middle columns). We will discuss in the next section that these different sequences, $1+3$ and $2+2$, correspond to different matching conditions for the phase differences in stationary correlators in order to construct LG strings for neutrino oscillations.

\section{Stationary Correlators}\label{ch2}

For the test of LGIs in neutrino oscillations, we consider situations where the flavour evolution can be approximated as that of a two-level system with an effective Hamiltonian that is independent of time. This is appropriate for neutrino oscillations in vacuum or in uniform matter as long as the observation times are much shorter than the oscillation period induced by the solar mass splitting. Under these conditions, the correlators take on the form:
\begin{multline}
\mathcal{C}_{ij} = p_+(t_i)[2P_{++}(t_j-t_i) - 1]\\ + p_-(t_i)[2P_{--}(t_j-t_i) -1]\,,
\end{multline}
where $p_\pm(t)$ is the probability of observing the system in the state $Q(t) = \pm1$ and $P_{\pm\pm}(\tau)$ is the conditional (``survival'') probability that the state is observed in the same state after a time difference $\tau$. If the evolution is invariant under time reversal, $P_{+-} = P_{-+}$, the survival probabilities become identical, $P_{++} = P_{--}$, and the correlators are independent of the system's initial condition with the simple form:
\begin{equation}\label{eq:Cnu}
\mathcal{C}_{ij} = \mathcal{C}(t_j-t_i)= 2P_{++}(t_j-t_i) - 1\,.
\end{equation}
We see that, under above conditions, the correlators $\mathcal{C}_{ij}$ become stationary, {\it i.e.}~only depend on the time between observations, and can be simply related to the stationary survival probability.

Consider now oscillation data obtained from an ensemble of neutrinos that are initialized in the (pure) state $Q = +1$ and observed at times $\tau_i>0$. Assuming stationarity as in Eq.~(\ref{eq:Cnu}), the LG strings defined in Eq.~(\ref{eq:Knsigma}) can be expressed in terms of $n$ time differences
\footnote{Since in both QM and MR we have $\mathcal{C}_{ij} = \mathcal{C}_{ji}$ we can focus on the absolute scale of time differences.} 
$\tau_i \equiv |t_{i+1}-t_i|$ and $\mathcal{C}_{i(i+1)} = \mathcal{C}(\tau_i)$. We can now consider a sequence ${\bf s} = (s_1,\ldots,s_n)$ representing the indices of $n$ time differences $\tau_{s_i}$ (including repeated entries) that can be written as a union of two sub-sequences, ${\bf s} = {\bf s}_a\cup {\bf s}_b$, with length $n_a$ and $n_b$ such that:
\begin{equation}\label{eq:phasematch}
\bigg|\sum_{i=1}^{n_a}\tau_{s_{a,i}} - \sum_{i=1}^{n_b}\tau_{s_{b,i}}\bigg|\leq  \frac{\varepsilon}{2}\sum_{i=1}^n\tau_{s_i}\,,
\end{equation}
where $\varepsilon \ll 1$ is the relative tolerance of time scale; to be discussed later. In the limit $\varepsilon \to 0$ the sequence ${\bf s}$ allows us to define a LG string:
\begin{equation}\label{eq:KSsigma}
K({\bf s},\boldsymbol\sigma) \equiv \sum_{i=1}^{n}\sigma_{i}\mathcal{C}(\tau_{s_i})\,,
\end{equation}
which is subject to the inequality $K({\bf s},\boldsymbol\sigma) \leq n-2$. For instance, Fig.~\ref{fig1} shows two examples of sequences for $n=4$ with $n_a=1$ and $n_b=3$ (``$1+3$''; middle column) and $n_a=2$ and $n_b=2$ (``$2+2$''; right column), where the time differences $\tau_i$ are equivalent to phase differences $\Delta\phi_i = \omega \tau_i$ in maximal two-level oscillations with some angular frequency $\omega$.

For the test of MR in neutrino oscillation, we can now determine the fraction of all sequences ${\bf s}$ that give at least one violation of LGIs. This is equivalent to testing the violation of the LGI for the maximum $K({\bf s},\boldsymbol\sigma)$ for a given sequence ${\bf s}$:
\begin{equation}\label{eq:Kmax}
K_{\rm max}({\bf s}) \equiv \max_{\boldsymbol\sigma} K({\bf s},\boldsymbol\sigma)\,,
\end{equation} 
analogous to the construction in Eq.~(\ref{eq:Knmax}). Note that the number of sequences satisfying the time-matching condition in Eq.~(\ref{eq:phasematch}) increases drastically with $n$. To avoid duplicates and trivial extensions of shorter sequences, {\it e.g.}~by simply adding the same time step to both sub-sequences, we also demand that: {\it a)} ${\bf s}$ is unique up to permutations, {\it b)} ${\bf s}_a$ and ${\bf s}_b$ are disjoint (${\bf s}_a \cap {\bf s}_b = \emptyset$), and {\it c)} ${\bf s}_a$ and ${\bf s}_b$ are not equivalent to unions ${\bf s}'_a\cup{\bf s}''_a$ and ${\bf s}'_b\cup{\bf s}''_b$, respectively, corresponding to shorter sequences ${\bf s}'$ and ${\bf s}''$ passing the time-matching condition.

\section{Macrorealistic Background}\label{ch3}

A challenging task for a statistical test of violation of LGIs is the definition of a ``realistic'' background hypothesis. Previous analyses of neutrino oscillation data approached this challenge by replacing the LG string in Eq.~(\ref{eq:Kn}) by a``classical'' LG string defined as~\cite{Formaggio:2016cuh,Fu:2017hky}:
\begin{equation}\label{eq:Kncl}
K^{\rm C}_n \equiv \sum_{i=1}^{n-1}\mathcal{C}_{i(i+1)} - \prod_{i=1}^{n-1}\mathcal{C}_{i(i+1)}\,.
\end{equation}
This expression is motivated by the assumption that a ``classical'' system behaves Markovian, which implies that the classical neutrino transition/survival probability $P^{\rm C}_{Q_iQ_j}$ obeys the Chapman-Kolmogorov equation~\cite{vanKampen:2007}:
\begin{equation}\label{eq:comp}
P^{\rm C}_{Q_1Q_3}(\tau_1+\tau_2) = \sum_{Q_2} P^{\rm C}_{Q_1Q_2}(\tau_1)P^{\rm C}_{Q_2Q_3}(\tau_2)\,.
\end{equation}
Assuming a two-level system obeying time-reversal symmetry, this implies that $\mathcal{C}(\tau_1+\tau_2) = \mathcal{C}(\tau_1)\mathcal{C}(\tau_2)$ and, more generally, $\mathcal{C}(\sum_i\tau_i) = \prod_{i}\mathcal{C}(\tau_i)$. 

The expression from Eq.~(\ref{eq:Kncl}) obeys the LGI $K^{\rm C}_n \leq n-2$ for arbitrary flavour correlations with a maximum of $K^{\rm C}_n$ appearing at $\mathcal{C}_{i(i+1)} = 1$. Therefore, regardless of the structure and precision of the neutrino oscillation data, the quantity from Eq.~(\ref{eq:Kncl}) will {\it never} violate the LGI and does not allow to quantify the level of chance violations in background data. The background distributions of LGI violations found in previous studies~\cite{Formaggio:2016cuh,Fu:2017hky} depend on the relatively rare cases where the neutrino oscillation data assumes {\it unphysical} values with $P>1$ or $P<0$ related to measurement uncertainties. This is the only case where the classical LG string in Eq.~(\ref{eq:Kncl}) can violate LGI. This demonstrates that an estimate of the statistical significance of LGI violations based on the classical LG string is not a robust method.

In this analysis we develop a more realistic approach. In general, we will assume that a macrorealistic system follows a classical Markovian process. The classical correlator of an (effective) two-level systems follows then the model $\mathcal{C}(\tau) = e^{-\Gamma\tau}$ where $\Gamma>0$ is the decorrelation rate; see also Ref.~\cite{Shafaq:2021lju}. We will consider two background models assuming: {\it a)} a fully correlated classical system with $\mathcal{C}(\tau) = 1$ assuming $\Gamma=0$, and {\it b)} a decorrelating system where the rate $\Gamma$ is fit to the experimental data. 
For each background model, denoted $\mathcal{H}_0^a$ and $\mathcal{H}_0^b$ in the following, we generate pseudo-data with the same variance $\Delta\mathcal{C}(\tau_i)$ as observed in the experimental data. In contrast to previous analyses, we will then determine the chance violations of LGIs based on statistical fluctuations of the macrorealistic background model.

\begin{figure}[t!]
\centering
\includegraphics[width=\linewidth]{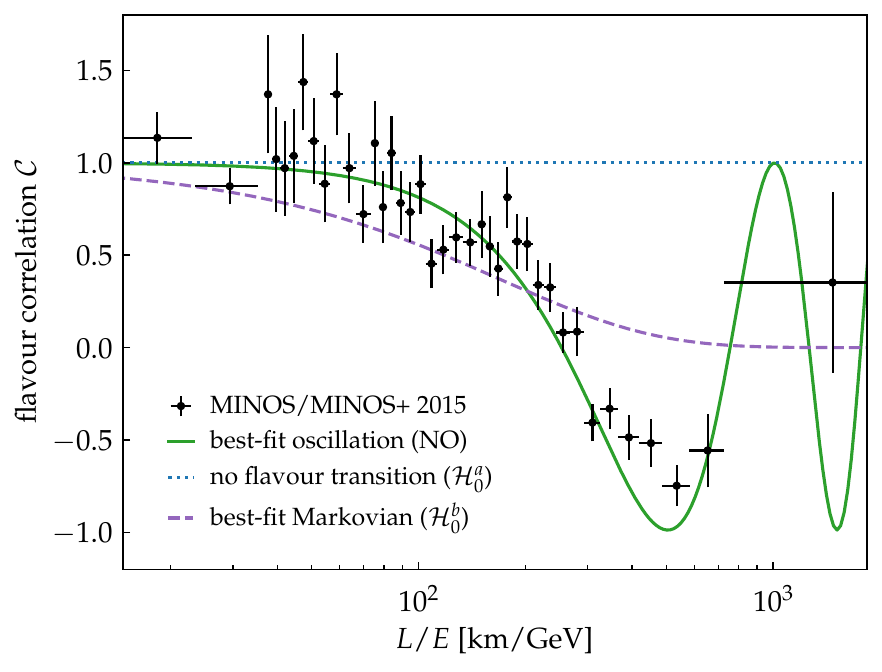}
\caption[]{Flavour correlations observed in MINOS/MINOS+ data~\cite{Sousa:2015bxa} (black data) in comparison to the prediction of the best-fit oscillation models assuming normal ordering~\cite{Esteban:2020cvm} (in green) and two classical models: a constant model (in dotted blue) and the best-fit Markovian model (in dashed purple).}\label{fig2}
\end{figure}

\section{Probing LGIs in Neutrino Data}\label{ch4}

In the following, we illustrate our revised test for violations of LGIs in neutrino oscillations using the muon neutrino survival probabilities reported by the MINOS experiment and its upgrade MINOS+~\cite{Sousa:2015bxa}. The oscillation data was inferred from muon neutrinos with energies up to $120$\,GeV produced by the NuMI beamline at Fermilab and observed as charged current events by near and far detectors over a baseline of $L\simeq735$~km. The MINOS/MINOS+ data in the form $P_{\rm obs}\pm\Delta P_{\rm obs}$ correspond to the ratio of observed $\nu_\mu$ events in the far detector compared to those expected from observations in the near detector and is provided in terms of 39 bins of reconstructed muon neutrino energy $E$~\cite{Sousa:2015bxa}. 

The observation of highly relativistic neutrinos at distance $L \simeq ct$ corresponds to neutrinos that have evolved over the proper time $t/\gamma \propto L/E$ since production, assuming that muon neutrinos have an (effective) mass $m\ll E$ and Lorentz factor $\gamma = E/m$. Hence, the role of time in neutrino data is played by the ratio $L/E$ and we will in general refer to this as the neutrino oscillation {\it phase}. Figure \ref{fig2} shows the MINOS/MINOS+ data in terms of the flavour correlations from Eq.~(\ref{eq:Cnu}) and $L/E$ bins. The solid green line shows the best-fit neutrino oscillation model using best-fit oscillations parameters assuming normal ordering~\cite{Esteban:2020cvm}. The baseline $L$ of the far detector is well below the oscillation length induced by the solar mass splitting, $\lambda_{\rm sol} \simeq 3.3\times10^4(E/{\rm GeV}){\rm km}$ and we can hence treat the oscillation as an effective two-level system driven by the oscillation length of the atmospheric mass splitting, $\lambda_{\rm atm} \simeq 990(E/{\rm GeV}){\rm km}$~\cite{Esteban:2020cvm}.

Assuming stationarity, we first determine the number of phase combinations ($\tau_i \propto L/E_i$) of the MINOS/MINOS+ data that allow us to construct LG strings. For comparison with the earlier study~\cite{Formaggio:2016cuh}, we choose our nominal phase tolerance in Eq.~(\ref{eq:phasematch}) to be $\varepsilon=0.5\%$. Table~\ref{tab1} shows the number of unique phase combinations following the prescription outlined in Section~\ref{ch2}. We show the results for up to $n=5$ combinations of phases in terms of their split into sub-sequences with length $n_a$ and $n_b$ (``$n_a+n_b$''). Note that the conventional LG strings in Eq.~(\ref{eq:Kn}) used in previous studies are based on the phase combinations with $n_a=1$ and $n_b=n-1$. However, as we pointed out in Section~\ref{ch1}, it is also possible to construct additional phase combinations in the case $n>3$ from the matching condition in Eq.~(\ref{eq:phasematch}).

\begin{table}[t!]
\centering\renewcommand{\arraystretch}{1.22}
\begin{ruledtabular}
\begin{tabular}{c|c|cc|cc}
\makebox[1.25cm][c]{$\varepsilon$} &
\makebox[1.25cm][c]{$n = 3$} &
\multicolumn{2}{c|}{\makebox[2.5cm][c]{$n = 4$}} &
\multicolumn{2}{c}{\makebox[2.5cm][c]{$n = 5$}}\\
&
\makebox[1.25cm][c]{$1+2$} & 
\makebox[1.25cm][c]{$1+3$} & 
\makebox[1.25cm][c]{$2+2$} & 
\makebox[1.25cm][c]{$1+4$} & 
\makebox[1.25cm][c]{$2+3$} \\
\hline
 5\%     & 731 & 6,902 & 9,479 & 45,990 & 218,542\\
 0.5\%   & 92  & 699   & 964   & 4,646  & 24,017\\
 0.05\%  & 24  & 76    & 81    & 495    & 2,350\\
\end{tabular}
\end{ruledtabular}
\caption[]{The total number of unique phase combinations satisfying Eq.~(\ref{eq:phasematch}) based on the MINOS/MINOS+ data. The combinations are shown for three values of the tolerance parameter $\varepsilon$ and increasing number of phases $n$, split into different combinations of sub-sequences ${\bf s}_a$ and ${\bf s}_b$ with $n = n_a + n_b$. Previous analyses of MINOS/MINOS+ data~\cite{Formaggio:2016cuh,Barrios:2023yub} are based on $1+2$ (in $K_3$) and $1+3$ (in $K_4$).}\label{tab1}
\end{table}

It is apparent in Fig.~\ref{fig2} that some of the correlations inferred from MINOS/MINOS+ data (black data) extend into the unphysical range $\mathcal{C}>1$, related to reconstructed survival probabilities with $P_{\rm obs}>1$. Note that the muon neutrino survival data reported by MINOS/MINOS+ correspond to ratios of observed muon neutrino events to those expected from background expectation assuming no flavour transitions. This quantity is subject to statistical fluctuations, reconstruction uncertainties, background subtraction, systematic uncertainties, {\it etc.}~and can lead to best-fit values of $P_{\rm obs}>1$ or even $P_{\rm obs}<0$. But the LGIs rely on the condition that the survival probabilities are bound to the physical region $0\leq P\leq 1$ (see Appendix~\ref{appA}). Therefore, in order to test for the violation of LGI it is necessary to \textit{account for unphysical correlations} related to measurement uncertainties.

\begin{figure*}[t!]
\includegraphics[height=0.39\linewidth]{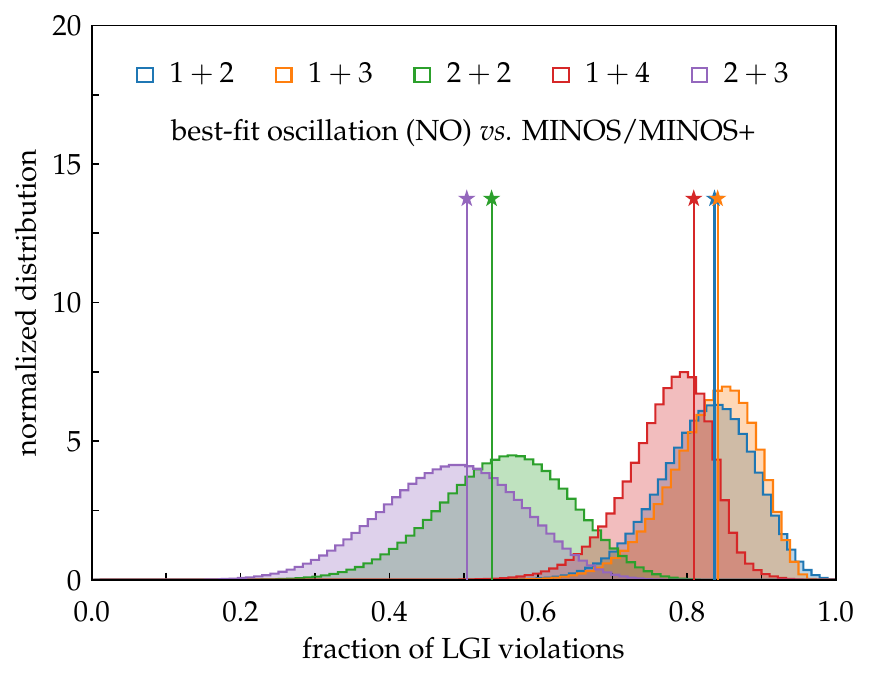}\hfill\includegraphics[height=0.39\linewidth]{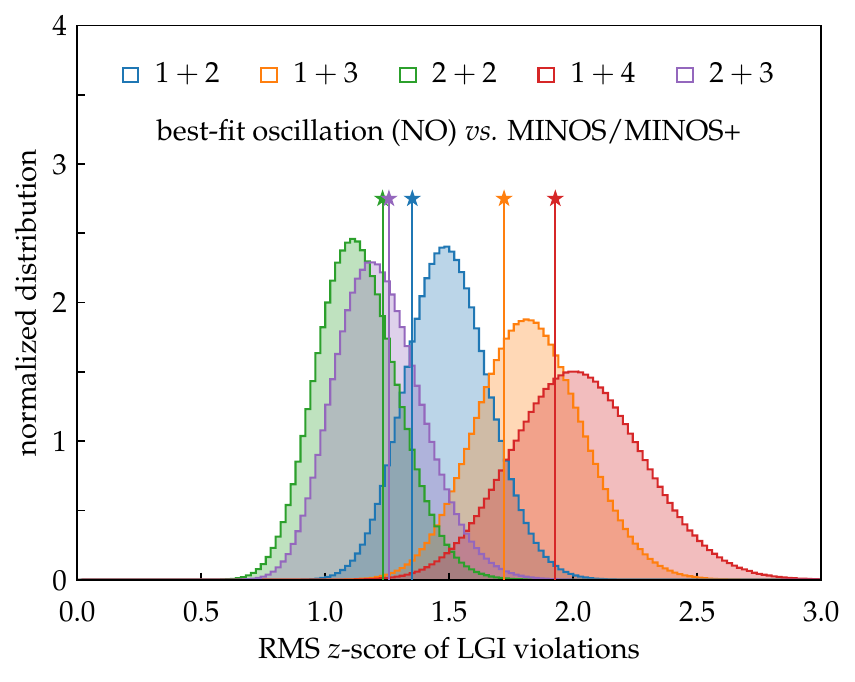}\\
\includegraphics[height=0.39\linewidth]{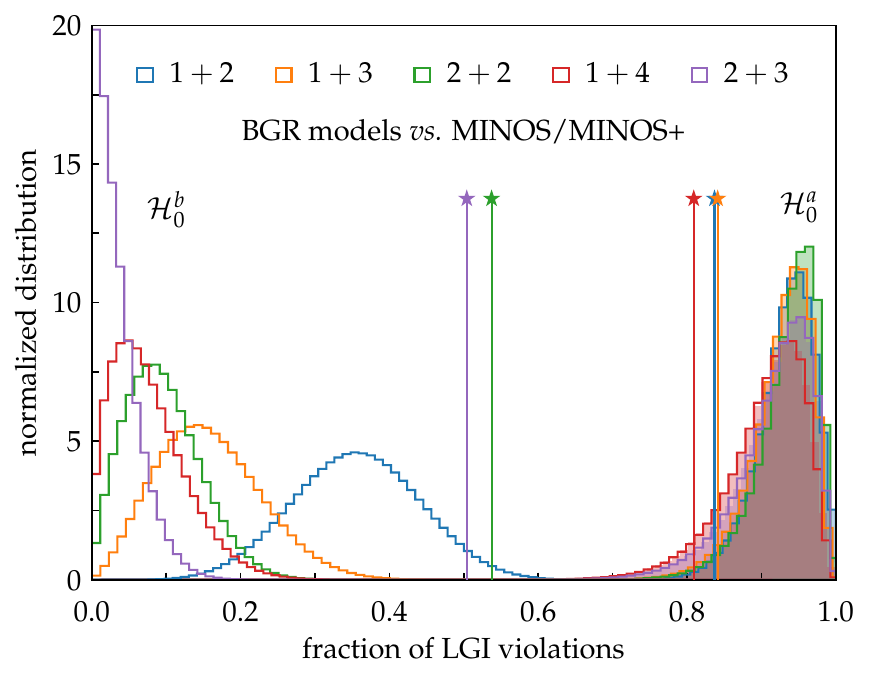}\hfill\includegraphics[height=0.39\linewidth]{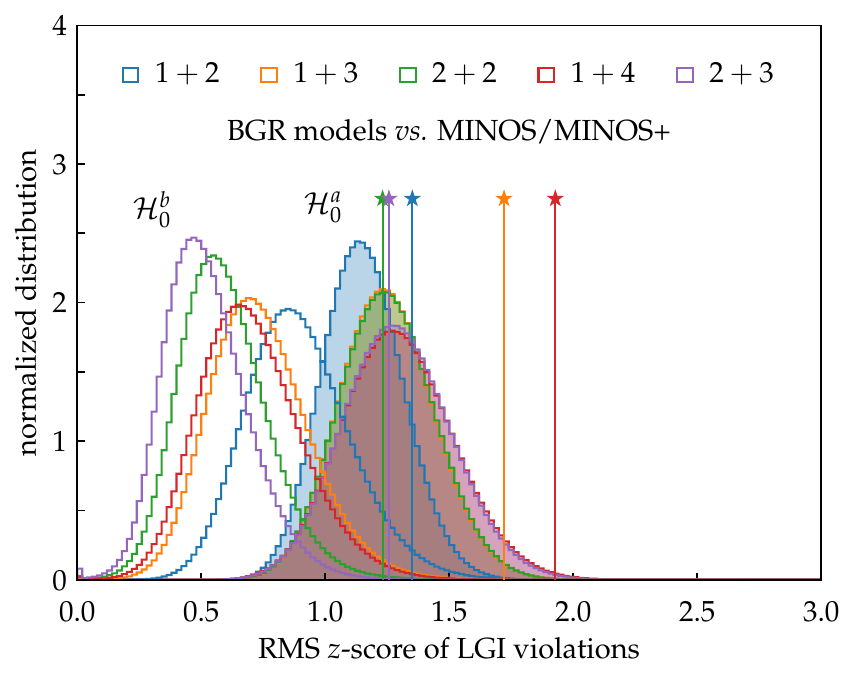}
\caption[]{Normalized distributions of the fraction of LGI violations (left panels) and corresponding RMS $z$-scores (right panels) based on $10^7$ pseudo-samples. Results are shown separately for LG strings in Eq.~(\ref{eq:Kmax}) for five different combinations of $n_a$ and $n_b$ with $n_a+n_b = 3$, $4$ and $5$ listed in Table~\ref{tab1}. The results from MINOS/MINOS+ data are shown as vertical lines with $\star$ symbols. The top panels show the distributions expected from neutrino oscillations assuming normal ordering (NO)~\cite{Esteban:2020cvm}. The bottom panels show those expected for the background (BGR) models $\mathcal{H}_0^a$ with $\mathcal{C}(\tau)=1$ (filled histograms) and $\mathcal{H}_0^b$ with $\mathcal{C}(\tau)=e^{-\Gamma \tau}$ (open histograms). The BGR distributions of RMS $z$-scores allow us to estimate the chance probabilities ($p$-values) of LGI violations under the BGR hypotheses at the level observed in MINOS/MINOS+ data. Treating the five $p$-values as independent trials, we estimate the post-trial significance at the level of $2.1\sigma$ ($3.7\sigma$) for the BGR model $\mathcal{H}_0^a$ ($\mathcal{H}_0^b$).}\label{fig3}
\end{figure*}

To quantify the level of LGI violations in the MINOS/MINOS+ data we now proceed as follows. We determine the fraction of LGI violations of the data for the five different phase combinations of LG strings listed in Table~\ref{tab1}. We then create pseudo-samples of data by sampling the muon neutrino survival probabilities from model predictions treating the (symmetric) measurement uncertainties $\Delta P_{\rm obs}$ of the MINOS/MINOS+ as the standard deviation of normal distributions. The $\phi \propto L/E$ bins of the pseudo-samples are identical to those of the MINOS/MINOS+ data and we average the model predictions over the bin width. As a first check, we can test if the data is consistent with the expectation from best-fit neutrino oscillation models assuming normal order~\cite{Esteban:2020cvm}, predicting a flavour correlation that is shown as the green line in Fig.~\ref{fig2}.

The top left plot in Fig.~\ref{fig3} shows the fraction of LGI violations observed in the MINOS/MINOS+ data (solid lines with $\star$ symbol) compared to the normalized distribution of those from $10^7$ pseudo-samples following the prediction of neutrino oscillation (filled histograms). The average fraction of LGI violations appears to be relatively high, reaching about 80\% on average for the case $n=3$ (blue histogram for $1+2$), consistent with Ref.~\cite{Formaggio:2016cuh}. The observed fraction of LGI violations lies well within the distributions of those from pseudo-samples in all case. However, to understand the significance of this result, we need to test the level of LGI violations that would be observed in macrorealistic data. This can be done by creating distributions from background expectations following our macrorealistic models, $\mathcal{H}_0^a$ and $\mathcal{H}_0^b$, introduced in the previous section and shown in terms of the corresponding flavour correlation in Fig.~\ref{fig2} (dotted and dashed lines, respectively).

The left bottom plot of Fig.~\ref{fig3} shows the fraction of LGI violations of the background model $\mathcal{H}_0^a$ with no transitions between neutrino flavours with $\mathcal{C}(\tau)=1$ (filled histograms) and the alternative background model $\mathcal{H}_0^b$ with $\mathcal{C}(\tau) = e^{-\Gamma\tau}$. One can observe that the expected LGI violations in the background models disagree with those of the oscillation model and, moreover, can drastically exceed those observed in the MINOS/MINOS+ data in the case of the model $\mathcal{H}_0^a$. The relatively large number of LGI violations for $\mathcal{H}_0^a$ can be understood from the definition of the LG string in Eq.~(\ref{eq:KSsigma}): The LGI is saturated for $\mathcal{C}=1$ and pseudo-data accounting for the MINOS/MINOS+ measurement uncertainties can frequently combine to lift the LG string above the boundary. In comparison, the chance violation of LGIs in the alternative background model $\mathcal{H}_0^b$ with $\mathcal{C}(\tau) = e^{-\Gamma\tau}$ are suppressed compared to those in the model $\mathcal{H}_0^a$ since the correlators are $|\mathcal{C}|\ll 1$ for most of the MINOS/MINOS+ data range.

The numerous LGI violations in the background model $\mathcal{H}_0^a$ indicate that the simple count of LGI violation is -- in general -- not a good statistical measure for tests of MR: small fluctuations in the pseudo-data derived for macrorealistic background models can frequently combine to violate LGIs. This motivates us to consider the (one-sided) $z$-score of a LGI violation for the phase combination ${\bf s}$ of length $n$ as: 
\begin{equation}\label{eq:zscore}
z({\bf s}) = {\rm max}\left(0,\frac{K_{\rm max}({\bf s})-n+2}{\Delta K_{\rm max}({\bf s})}\right)\,,
\end{equation}
where $K_{\rm max}({\bf s})$ is the maximum LG string in Eq.~(\ref{eq:Kmax}) and $\Delta K_{\rm max}({\bf s},\boldsymbol\sigma)$ its uncertainty following from the uncertainty of the MINOS/MINOS+ data via error propagation. 
The root-mean-square (RMS) of the $z$-scores for a given pseudo-sample with $N_{\rm LGV}>0$ defines our test statistic (TS):
\begin{equation}\label{eq:TS}
z_{\rm RMS} \equiv \sqrt{\frac{1}{N_{\rm LGV}}\sum_{{\bf s}}z^2({\bf s})}\,,
\end{equation}
where $N_{\rm LGV}$ are the number of sequences that {\it violate} the LGI, {\it i.e.}~sequences with $z({\bf s})>0$. In the case $N_{\rm LGV} = 0$ we instead use $z_{\rm RMS} = 0$.

The resulting distributions of RMS $z$-scores of LGI violations are shown in the right column of Fig.~\ref{fig3}. The values derived from the MINOS/MINOS+ data (vertical lines with $\star$ symbols) show an excess of $z_{\rm RMS}>1$ for all five combinations of LG sequences, most noticeably for the cases $1+3$ and $1+4$ and consistent with the expected TS distribution from the best-fit oscillation model (top right panel). The bottom right panel shows the corresponding TS distribution of the two background models. Despite the numerous violations of LGIs in the background case $\mathcal{H}_0^a$ (filled histograms in bottom left panel), the corresponding RMS $z$-scores only marginally exceed unity (filled histograms in bottom right panel). Note that the TS distributions of this time-independent background model ($\mathcal{C}=1$) depend on the total number of phases, $n=n_a+n_b$, rather than the individual split in $n_a$ and $n_b$. This is noticeable as the near-identical distributions for $1+3$ and $2+2$ as well as $1+4$ and $2+3$. On the other hand, the TS distributions of the background case $\mathcal{H}_0^b$ (open histograms) are centered around much lower $z_{\rm RMS}$ values and those of the MINOS/MINOS+ data appear in their high-TS tails.

As we discuss in Appendix~\ref{appC}, our TS in Eq.~(\ref{eq:TS}) is related to a maximum log-likelihood ratio of the set of LG strings $K_{\rm max}({\bf s})$, assuming that their true value is limited by the boundary $n-2$ under the background hypothesis. The RMS $z$-scores are then equivalent to the reduced $\chi^2$ where the degree of freedom (dof) corresponds to the number of LGI violations $N_{\rm LGV}$. For LGI violations driven by quantum mechanics (and not by measurement uncertainties), we expect that the reduced $\chi^2$ exceeds $1$. This is clearly noticeable in the TS distributions of the oscillation model and the MINOS/MINOS+ data. However, also the TS distributions for the background model $\mathcal{H}_0^a$ have a median that marginally exceed $1$, related to the optimization of the sign vector $\boldsymbol\sigma$ in Eq.~(\ref{eq:Kmax}) introducing a bias towards larger $z$-scores above the boundary $n-2$.

\section{Discussion}\label{ch5}

The RMS $z$-score from Eq.~(\ref{eq:TS}) allows us to estimate the significance of LGI violations in neutrino oscillation data. For the TS distribution of individual families of LG sequences ``$n_a+n_b$'', the $p$-value of the violation of LGI can be determined by the fraction of background pseudo-samples with a TS larger than the TS observed for the actual data, as shown in the bottom right panel of Fig.~\ref{fig3}. The five $p$-values that we determine this way for each background model can -- in general -- not be considered as independent trials, since they are based on the same pseudo-samples, each consisting of only 39 data points. Conservatively, we estimate the post-trial $p$-value from the minimum $p$-value $p_{\rm min}$ as $p_{\rm post} = 1-(1-p_{\rm min})^5$. This yields a post-trial significance of LGI violations of $2.1\sigma$ ($3.7\sigma$) in comparison to the background model $\mathcal{H}_0^a$ ($\mathcal{H}_0^b$) with $p_{\rm min} \simeq 2.0\times10^{-2}$ ($8.9\times10^{-5}$).

We argue that the background case $\mathcal{H}_0^a$ corresponds to the {\it most robust estimate} for the significance of LGI violations in neutrino oscillations, reflected by the smaller post-trial significance. A background expectation value $\mathcal{C}=1$ guarantees that the maximal LG string $K_{\rm max}$ falls closed to the boundary $n-2$ of LGI for any sequence ${\bf s}$. Consequently, the pseudo-samples derived from this background model allow us to directly estimate the impact of measurement uncertainties on the chance violation of LGIs. On the other hand, the background case $\mathcal{H}_0^b$ with exponentially decaying correlations has a lower chance for LGI violations, as visible in the bottom panels of Fig.~\ref{fig3}. In particular, for larger $n$ the chance of violation of LGIs in the background cases reduces significantly. 

It is important to emphasize that a test of LGI violation in neutrino flavour evolution is not equivalent to a test of neutrino oscillations {\it per se}. A constant flavour correlation $\mathcal{C}=1$ is not consistent with the MINOS/MINOS+ data in Fig.~\ref{fig2}. Describing the goodness of fit of the three models shown in Fig.~\ref{fig2} to the original MINOS/MINOS+ data by $\chi^2$ values, we see that the best-fit Markovian model with $\chi^2/\mathrm{dof} \simeq 5.8$ is a significantly better description of the MINOS/MINOS+ data than the constant model, which has $\chi^2/\mathrm{dof} \simeq 31$. The best-fit oscillation model assuming normal ordering has $\chi^2/\mathrm{dof} \simeq 1.8$. These different goodness-of-fit values are also reflected by the different number $N_{\rm LGV}$ of LGI violations of these models (left column of Fig.~\ref{fig3}). The normalization of the TS of Eq.~(\ref{eq:TS}) via $N_{\rm LGV}$ allows us to focus on the {\it quality} of the LGI violation, not their {\it quantity.}

Before concluding, we would also like to comment on the phase tolerance $\varepsilon$ that determines the number of sequences $\bf s$ of LG strings. For comparison to earlier studies we choose $\varepsilon = 0.5\%$ as our benchmark value. The flavour correlation data is effectively averaged over a time scale corresponding to the size of the energy bins and this sets an upper limit for the required resolution of the phase matching. The relative energy uncertainty $\Delta E/E$ of the MINOS/MINOS+ data reaches 3--6\% in the 2--20\,eV energy region, so the choice of $\varepsilon = 0.5\%$ is conservative. Switching to $\varepsilon = 5\%$ with a larger number of phase combinations (see Table~\ref{tab1}) leaves our results practically unchanged.

\section{Conclusions}\label{ch6}

We discussed a test of Leggett-Garg inequalities (LGIs) in neutrino flavour oscillations. These inequalities relate the correlation of consecutive measurements in an evolving system and are analogous to Bell's inequalities of measurements performed on spatially separated systems. The violation of LGIs in neutrino oscillations serves as a test of the assumption of {\it macrorealism} in the broader sense and constitutes a fundamental test of the quantum nature of neutrinos.

Our work improves the methodology of previous analyses of LGIs in several ways. We derived a generalized family of Leggett-Garg strings that lead to an optimal test for the violation of LGIs for a given set of neutrino flavour measurements. We provided an improved method to simulate pseudo-data following a statistical background hypothesis of macrorealistic systems, which is not biased towards unphysical flavour correlations and their corresponding spurious violations of LGIs. We introduced a test statistic that allows us to give a robust estimate of the significance of LGI violations in neutrino data.

We illustrated our revised methodology based on muon neutrino survival data by the MINOS detector. Using our method we find that MINOS/MINOS+ data~\cite{Sousa:2015bxa} violates LGI with a significance at the level of $(2-3)\sigma$, depending on background model. 
The main reason for the reduced significance compared to the earlier claim of $6.2 \sigma$ of Ref.~\cite{Formaggio:2016cuh} comes from our frequentist approach to quantify the level of LGI violation on data following a background hypothesis, rather than introducing a classical LG string as a revised test statistic. Other work based on accelerator and reactor neutrino data has either used the same background method~\cite{Fu:2017hky} or neglected a background model altogether~\cite{Barrios:2023yub,Wang:2022tnr}. Our findings indicate that also these results should be critically re-examined with respect to the statistical level of LGI violations.

\vspace*{0.1cm}

\begin{acknowledgments}
M.A., K.M.G.~and J.I.-N.~acknowledge support by Villum Fonden (No.~18994). D.J.K.~acknowledges support from the Carlsberg Foundation (No.~117238). J.I.-N. performed part of this work as a visiting researcher supported by the Erasmus+ program of the European Union. 
\end{acknowledgments}

\bibliographystyle{utphys_mod}
\bibliography{references}

\providecommand{\href}[2]{#2}\begingroup\raggedright\begin{thebibliography}{10}

\bibitem{Bell:1964kc}
J.~S. Bell, \href{http://dx.doi.org/10.1103/PhysicsPhysiqueFizika.1.195}{{\em
  Physics Physique Fizika} {\bfseries 1} (1964) 195--200}.

\bibitem{Leggett:1985zz}
A.~J. Leggett and A.~Garg,
  \href{http://dx.doi.org/10.1103/PhysRevLett.54.857}{{\em Phys. Rev. Lett.}
  {\bfseries 54} (1985) 857--860}.

\bibitem{Leggett:2002}
A.~J. Leggett, \href{http://dx.doi.org/10.1088/0953-8984/14/15/201}{{\em J.
  Phys.: Condens. Matter} {\bfseries 14} (2002) R415}.

\bibitem{Emary:2013wfl}
C.~Emary, N.~Lambert, and F.~Nori,
  \href{http://dx.doi.org/10.1088/0034-4885/77/1/016001}{{\em Rept. Prog.
  Phys.} {\bfseries 77} no.~1, (2013) 016001}.

\bibitem{Gangopadhyay:2013aha}
D.~Gangopadhyay, D.~Home, and A.~S. Roy,
  \href{http://dx.doi.org/10.1103/PhysRevA.88.022115}{{\em Phys. Rev. A}
  {\bfseries 88} no.~2, (2013) 022115},
  \href{http://arxiv.org/abs/1304.2761}{{\ttfamily arXiv:1304.2761}}.

\bibitem{Gangopadhyay:2017nsn}
D.~Gangopadhyay and A.~S. Roy,
  \href{http://dx.doi.org/10.1140/epjc/s10052-017-4837-2}{{\em Eur. Phys. J. C}
  {\bfseries 77} no.~4, (2017) 260},
  \href{http://arxiv.org/abs/1702.04646}{{\ttfamily arXiv:1702.04646}}.

\bibitem{Formaggio:2016cuh}
J.~A. Formaggio, D.~I. Kaiser, M.~M. Murskyj, and T.~E. Weiss,
  \href{http://dx.doi.org/10.1103/PhysRevLett.117.050402}{{\em Phys. Rev.
  Lett.} {\bfseries 117} no.~5, (2016) 050402},
  \href{http://arxiv.org/abs/1602.00041}{{\ttfamily arXiv:1602.00041}}.

\bibitem{Barrios:2023yub}
R.~Z. Barrios and M.~A. Acero,
  \href{http://arxiv.org/abs/2401.00240}{{\ttfamily arXiv:2401.00240}}.

\bibitem{Fu:2017hky}
Q.~Fu and X.~Chen, \href{http://dx.doi.org/10.1140/epjc/s10052-017-5371-y}{{\em
  Eur. Phys. J. C} {\bfseries 77} no.~11, (2017) 775},
  \href{http://arxiv.org/abs/1705.08601}{{\ttfamily arXiv:1705.08601}}.

\bibitem{Wang:2022tnr}
X.-Z. Wang and B.-Q. Ma,
  \href{http://dx.doi.org/10.1140/epjc/s10052-022-10053-1}{{\em Eur. Phys. J.
  C} {\bfseries 82} no.~2, (2022) 133},
  \href{http://arxiv.org/abs/2201.10597}{{\ttfamily arXiv:2201.10597}}.

\bibitem{Song:2018bma}
X.-K. Song, Y.~Huang, J.~Ling, and M.-H. Yung,
  \href{http://dx.doi.org/10.1103/PhysRevA.98.050302}{{\em Phys. Rev. A}
  {\bfseries 98} no.~5, (2018) 050302},
  \href{http://arxiv.org/abs/1806.00715}{{\ttfamily arXiv:1806.00715}}.

\bibitem{Naikoo:2019eec}
J.~Naikoo, A.~Kumar~Alok, S.~Banerjee, and S.~Uma~Sankar,
  \href{http://dx.doi.org/10.1103/PhysRevD.99.095001}{{\em Phys. Rev. D}
  {\bfseries 99} no.~9, (2019) 095001},
  \href{http://arxiv.org/abs/1901.10859}{{\ttfamily arXiv:1901.10859}}.

\bibitem{Shafaq:2021lju}
S.~Shafaq, T.~Kushwaha, and P.~Mehta,
  \href{http://arxiv.org/abs/2112.12726}{{\ttfamily arXiv:2112.12726}}.

\bibitem{vanKampen:2007}
N.~G. van Kampen, {\em {Stochastic Processes in Physics and Chemistry}}.
\newblock Elsevier, 2007.

\bibitem{Sousa:2015bxa}
A.~B. Sousa, (MINOS, MINOS+  Collaboration),
  \href{http://dx.doi.org/10.1063/1.4915576}{{\em AIP Conf. Proc.} {\bfseries
  1666} no.~1, (2015) 110004},
  \href{http://arxiv.org/abs/1502.07715}{{\ttfamily arXiv:1502.07715}}.

\bibitem{Esteban:2020cvm}
I.~Esteban, M.~C. Gonzalez-Garcia, M.~Maltoni, T.~Schwetz, and A.~Zhou,
  \href{http://dx.doi.org/10.1007/JHEP09(2020)178}{{\em JHEP} {\bfseries 09}
  (2020) 178}, \href{http://arxiv.org/abs/2007.14792}{{\ttfamily
  arXiv:2007.14792}}. \href{http://www.nu-fit.org}{\tt http://www.nu-fit.org}.

\end{thebibliography}\endgroup

\appendix

\section{Sequential Leggett-Garg Inequalities}\label{appA}

By definition, MR assumes that the $n$ non-invasive measurements $Q_i$ follow a probability distribution $\rho(Q_1,\ldots,Q_n)$. Since $Q_i = \pm1$, we can label each of the $2^n$ possible outcomes by an integer: 
\begin{equation}\label{eq:ell1}
\ell = \sum_{i=1}^{n} 2^{i-1}\frac{1+\sigma_iQ_i}{2}\,,
\end{equation}
where $\sigma_i=\pm1$ is an arbitrary sign convention for {\it each} measurement $Q_i$.
The terms $(1+\sigma_iQ_i)/2 \in \lbrace0,1\rbrace$ are the bits $\ell_{i-1}$ in the binary representation of $\ell$. The macrorealistic correlator $\mathcal{C}_{ij}$ from Eq.~(\ref{eq:CMR}), of the measurements $Q_i$ and $Q_j$ is then related to the $2^n$ probabilities $P_\ell$ of observing the $\ell$th configuration as:
\begin{equation}\label{eq:Cconst}
 \sigma_i\sigma_j\mathcal{C}_{ij} = \sum_{\ell=0}^{2^n-1}(-1)^{\ell_{i-1}+\ell_{j-1}} P_\ell\,,
\end{equation}
with $\ell_i = \lfloor{\ell}/{2^i}\rfloor\!\!\!\mod 2$. 

We first start with the usual sign convention $\sigma_i=\pm1$. For three observables $Q_1$, $Q_2$, and $Q_3$ the explicit expressions of the three different correlators are:
\begin{align}
\mathcal{C}_{12} &= P_0 - P_1 - P_2 + P_3 + P_4 - P_5 - P_6 + P_7\,,\\
\mathcal{C}_{23} &= P_0 + P_1 - P_2 - P_3 - P_4 - P_5 + P_6 + P_7\,,\\
\mathcal{C}_{13} &= P_0 - P_1 + P_2 - P_3 - P_4 + P_5 - P_6 + P_7\,.
\end{align}
The LG string $K_3$ defined in Eq.~(\ref{eq:Kn}) is now:
\begin{equation}
K_3 = \mathcal{C}_{12} + \mathcal{C}_{23} - \mathcal{C}_{13} = 1 - 4(P_2+P_5) \leq 1\,,
\end{equation}
where we have used $\sum_\ell P_\ell = 1$ in the last step.

Consider now the $n^{\rm th}$ LG string for $n>3$. We can relate $K_n$ to the previous string $K_{n-1}$ via:
\begin{equation}
K_{n} =  K_{n-1}  + \mathcal{C}_{1n-1} + \mathcal{C}_{n-1n} - \mathcal{C}_{1n}\,.
\end{equation}
Using the LGI for $K_3$ we know that:
\begin{equation}
 \mathcal{C}_{1n-1} + \mathcal{C}_{n-1n} - \mathcal{C}_{1n} \leq 1\,.
\end{equation}
We therefore arrive at the LGI for the $n$th LG string:
\begin{equation}
K_{n} \leq K_{n-1} +1 \leq n-2\,,
\end{equation}
where we have used the LGI for $K_{n-1}$ in the last step.

Finally, instead of the sign convention $\sigma_i = 1$ that we used in the preceding derivation, we would arrive at the same result by replacing the classical correlators in the LG string by $\sigma_i\sigma_j\mathcal{C}_{ij}$ which appears on the l.h.s.\@ of Eq.~(\ref{eq:Cconst}). Since the sign vector $\boldsymbol\sigma$ yields the same expression as the reverse $-\boldsymbol\sigma$ we can focus on the number of sign combinations with $\sigma_1=1$. If we now define $\sigma'_i \equiv \sigma_{i}\sigma_{i+1}$ we arrive at the generalized LG strings of Eq.~(\ref{eq:Knsigma}) where the signs are arbitrary but need to obey the relation in Eq.~(\ref{eq:sigmacon}).

\section{Violation of Leggett-Garg Inequalities}\label{appB}

Consider the (pure) neutrino state $|\psi\rangle$ that is initially
($t=0$) in an arbitrary superposition of flavour states. A first measurement at time $t_1$ can yield a neutrino of flavour state $\nu_\mu$ ($|+\rangle$) or either $\nu_e$ or $\nu_\tau$ ($|-\rangle$). A second measurement at $t_2>t_1$ yields one of the two flavour groups ($\pm'$). According to the Born rule, the probabilities of measuring the four flavour group combinations at times $t_1$ and $t_2$ are:
\begin{equation}\label{eq:Pprob}
P(\pm,\pm') = \langle \psi|\hat\pi_\pm(t_1)\hat\pi_{\pm'}(t_2)\hat\pi_\pm(t_1)|\psi\rangle\,,
\end{equation}
where we define the Heisenberg operator $\hat\pi_\pm(t) = U(t)^\dagger \hat\pi_\pm U(t)$ with time evolution operator $U(t)$ and projection operator $\hat\pi_\pm \equiv ( 1\pm \hat Q)/2$. The quantum-mechanical correlator at times $t_1$ and $t_2$ can then be written as:
\begin{align}\label{eq:C12}
\mathcal{C}_{12}  &= P(+,+) + P(-,-) - P(+,-) - P(-,+)\,,
\end{align}
which is equivalent to Eq.~(\ref{eq:CQM}) with $\hat Q_i \equiv U^\dagger(t_i)\hat Q U(t_i)$. We can generalize this to a mixed ensemble of states as $\mathcal{C}_{12} = {\rm Tr}(\lbrace\hat Q_1,\hat Q_2\rbrace\rho)/2$ with density matrix $\rho$.

The time-evolution of $\hat Q$ is equivalent to a rotation of a Pauli spinor.
A general time-independent two-flavour Hamiltonian, {\it e.g.}~neglecting spatial variations of the matter potential during neutrino propagation, takes the form $\hat H = (a + {\bf b}\cdot\boldsymbol{\sigma})/2$ where $\boldsymbol\sigma$ represents a vector of Pauli matrices and $a$ ($\bf b$) a real scalar (vector) with dimensions of energy. From this, we obtain:
\begin{equation}
\hat Q(t) = ({\bf n}\cdot\boldsymbol\sigma)n_z(1-\cos\phi) + \cos\phi\sigma_z + \sin\phi({\bf n}\times\boldsymbol\sigma)_z\,,
\end{equation}
with ${\bf n} = {\bf b}/|{\bf b}|$ and $\phi = t|{\bf b}|$. The correlator in Eq.~(\ref{eq:C12}) is then:
\begin{equation}
\mathcal{C}_{12} = n_z^2+(1-n_z^2)\cos(\phi_1-\phi_2)\,.
\end{equation}
For vacuum oscillations with mixing $|\nu_1\rangle = \cos\theta|+\rangle + \sin\theta|-\rangle$ and $|\nu_2\rangle = \cos\theta|-\rangle - \sin\theta|+\rangle$ between two effective mass states $|\nu_i\rangle$ with mass splitting $\Delta m^2$ and energy $E_\nu$ we have ${\bf b} = \omega\left(\sin2\theta,0,\cos2\theta\right)$
with oscillation frequency $\omega = |{\bf b}| = \Delta m^2/2E_\nu$ and:
\begin{equation}
\mathcal{C}_{12} = \cos^22\theta+\sin^22\theta\cos\left(\omega(t_2-t_1)
\right)\,.
\end{equation}
The maximal violation of LGIs can be observed under maximal mixing, {\it i.e.}~mixing angle $\theta = \pi/4$.

\section{Boundary Test Statistic}\label{appC}

We motivate our test statistic in Eq.~(\ref{eq:TS}) as follows. Assuming that $n$ data $x_i$ are normal distributed with standard deviation $\Delta x_i$ and expectation values $\mu_i$, we define the likelihood as the product of normal distributions:
\begin{equation}
\mathcal{L}(\boldsymbol\mu|{\bf x}) = \prod_{i=1}^n\frac{1}{\sqrt{2\pi}\Delta x_i}\exp\left(-\frac{(x_i-\mu_i)^2}{2(\Delta x_i)^2}\right)\,.
\end{equation}
In the following we are agnostic about the exact model predictions $\mu_i$. However, under our background hypothesis $\mathcal{H}_0$ we assume that the background model is bounded to values $\mu_i \leq B$. Treating the $\mu_i$ as independent parameters, we can maximize the likelihood for the case of background ($\mathcal{H}_0$) and signal ($\mathcal{H}$) hypotheses as $\widehat{\mu}_i = x_i$ and $\widehat{\mu}_{0,i} = \min(B,x_i)$, respectively. The maximum log-likelihood-ratio is then:
\begin{equation}\label{eq:t}
t\equiv-2\ln\frac{\mathcal{L}(\widehat{\boldsymbol\mu}_0|{\bf x})}{\mathcal{L}(\widehat{\boldsymbol\mu}|{\bf x})} = \sum_{i=1}^n z_i^2\,,
\end{equation}
with one-sided $z$-scores defined as:
\begin{equation}\label{eq:zorig}
z_i \equiv \max\left(0,\frac{x_i-B}{\Delta x_i}\right)\,,
\end{equation}
which reproduces Eq.~(\ref{eq:zscore}) after the replacements $B\to n-2$ and $x_i \to K_{\rm max}({\bf s})$.

For background data that follow normal distributions with mean $\mu_i = B$ we expect that Eq.~(\ref{eq:t}) follows a $\chi^2$-distribution with degree of freedom corresponding to the number $N_{\rm V}\leq n$ of boundary violations, $x_i>B$. The number of violations $N_{\rm V}$ depends not only on the signal and background hypotheses, but also on the variance $(\Delta x_i)^2$ related to measurement uncertainties. This motivates us to define a test statistic as:
\begin{equation}\label{eq:tV}
t_{\rm V} \equiv \begin{cases}t/N_{\rm V}& N_{\rm V}>0\,,\\
0 & N_{\rm V}=0\,,\end{cases}
\end{equation}
equivalent to a reduced $\chi^2$ and identical to the square of the RMS $z$-score in Eq.~(\ref{eq:TS}). For general background data sampled from expectation values $\mu_i \leq B$ the median $t_{\rm V}$ value will lie below $1$. On the other hand, for data sampled from signal models where $\mu_i$ can (but not must) exceed the boundary, the median $t_{\rm V}$ value can significantly exceed $1$. 

The background model $\mathcal{H}_0^a$ introduced in Section~\ref{ch3} with $\mathcal{C} = 1$ produces LG strings that are expected to scatter around the boundary $B=n-2$. We therefore consider this model as the most conservative estimate for the TS distribution of data following macrorealistic background hypotheses. Note that we determine the significance of LGI violations from the TS distributions of pseudo-samples following the background and signal hypotheses. This construction does not rely on detailed statistical properties of Eq.~(\ref{eq:tV}), but rather on its tendency to increase beyond $1$ in the presence of signal regardless of the overall number of violations.

\end{document}